\documentstyle[11pt,newpasp,twoside]{article}
\def\edcomment#1{\iffalse\marginpar{\raggedright\sl#1\/}\else\relax\fi}
\reversemarginpar

\begin{document}
\title{Combining Astrometry and Spectroscopy}
 \author{Guillermo Torres}
\affil{Harvard-Smithsonian Center for Astrophysics, 60 Garden St.\
Mail Stop 20, Cambridge, MA 02138, USA}

\begin{abstract} 
 Orbital solutions for binary or multiple stellar systems that combine
astrometry (e.g., position angles and angular separations) with
spectroscopy (radial velocities) have important advantages over
astrom\-etr\-ic-only or spectroscopic-only solutions. In many cases
they allow the determination of the absolute masses of the components,
as well as the distance. Yet, these kinds of combined solutions that
use different types of observations in a global least-squares fit are
still not very common in the literature. An outline of the procedure
is presented, along with examples to illustrate the sort of results
that can be obtained. The same method can easily be extended to
include other types of measurements (times of eclipse, Hipparcos
observations, interferometric visibilities and closure phases,
parallaxes, lunar occultations, etc.), which often complement each
other and strengthen the solution. 
 \end{abstract}

\section{Introduction}

The main subject of this review is orbital solutions in binary and
multiple stellar systems, with emphasis on bringing together
observations from different techniques (mainly astrometry and
spectroscopy) into a single fit that is usually better than separate
solutions using either of those kinds of data. Orbit determination is
a classical discipline in astronomy that dates back more than 170
years. The first astrometric orbit determination, based on
measurements of the angular separation and position angle of the
visual binary $\xi$~UMa, is credited to F\'elix Savary (1827).  The
first determination of a spectroscopic orbit was made by Arthur A.\
Rambaut (1891), based on radial velocity measurements of $\beta$~Aur
by E.\ C.\ Pickering. Since then numerous algorithms have been
developed for both kinds of solutions, which are well described in
textbooks on the subject, and many thousands of orbits have been
determined. 

Because astrometric measurements describe motion on the plane of the
sky, and spectroscopic measurements (radial velocities) describe
motion along the line of sight, the two techniques are complementary
and it is fairly obvious that there are advantages in combining them
for the same system. This can easily be seen by looking at the
classical orbital elements used to describe each type of orbit.  For
astrometric orbits of binaries the elements are $P$, $a\arcsec$, $e$ ,
$i$, $\omega$, $\Omega$, and $T$, which have their usual meaning, and
where we indicate by $a\arcsec$ the \emph{angular} semimajor axis, to
distinguish it from the \emph{linear} semimajor axis that appears as a
derived quantity in spectroscopy (see below).  The conventional
spectroscopic elements for a double-lined spectroscopic binary are
$P$, $\gamma$, $K_1$, $K_2$, $e$, $\omega$, and $T$, with the usual
caveat that $\omega$ here is the longitude of periastron for the
primary component, whereas the normal convention in visual orbits is
to use the longitude of periastron for the secondary component.
Trivially the two angles differ by $180\deg$. Four orbital elements
are in common between astrometric and spectroscopic solutions: $P$,
$e$, $\omega$, and $T$, which means that both kinds of observations
can constrain them. 

\section{Combined solutions: Advantages and disadvantages}

Having astrometric and spectroscopic observations available for the
same system allows a much more complete description of the path of the
stars in space (sometimes referred to as a ``three-dimensional
orbit").  In addition, new information can be obtained by combining
the two kinds of data, most notably the individual masses of the
components (in double-lined astrometric-spectroscopic binaries), and
the ``orbital parallax". The latter allows a direct determination of
the luminosities of the stars. 

The absolute masses in a binary follow from the expressions for the
minimum mass derived from the spectroscopic solution,
 \begin{eqnarray*}
M_1 \sin^3 i & = & P (1-e^2)^{3/2}(K_1+K_2)^2 K_2 \\
M_2 \sin^3 i & = & P (1-e^2)^{3/2}(K_1+K_2)^2 K_1, 
\end{eqnarray*}
 and the inclination angle $i$ provided by astrometry. 

The orbital parallax is a direct measure of the distance free from any
assumptions beyond Newtonian physics, and can often be more precise
than trigonometric parallaxes, particularly for more distant systems.
It follows trivially from the ratio between the projected angular
semimajor axis from astrometry ($a\arcsec \sin i$), and the projected
linear semimajor axis from spectroscopy,
 \begin{displaymath}
a \sin i = P (K_1+K_2)\sqrt{1-e^2},
\end{displaymath}
 as
\begin{displaymath}
\pi_{\rm orb} = {a\arcsec \sin i\over P (K_1+K_2)\sqrt{1-e^2}}.
\end{displaymath}

If we are interested in combining the two different types of data, the
question then arises as to how to do this in the optimal way.  One
possibility, which is by far the most common approach seen in the
literature, is simply to carry out completely separate astrometric and
spectroscopic fits if possible, and proceed to use the information
provided by these separate solutions to derive other properties of the
system (such as $M_1$, $M_2$, and $\pi_{\rm orb}$).  In this approach
the four elements in common between the two types of solutions ($P$,
$e$, $\omega$, and $T$) are usually averaged, possibly with some
weighting, whereas the elements $a\arcsec$, $i$, and $\Omega$ are
adopted from the visual solution, and $\gamma$, $K_1$ and $K_2$ are
taken directly from the spectroscopy. One disadvantage of this
approach is that separate spectroscopic and astrometric solutions are
not always possible for a given system. This can happen, for instance,
when the phase coverage of one type of observation is insufficient for
an independent fit, even though the other type of observation may
allow a good solution.  In addition, by averaging the four elements in
common, the uncertainties to be assigned to those elements are
ill-defined and often not very realistic. Separate solutions ignore
the redundancy provided by the measurements, and are wasteful of
information that is sometimes critical to avoid systematic errors. 

The second approach to a combined astrometric-spectroscopic solution
is to merge all the data into a single least-squares fit, instead of
having two separate fits. In this case one solves simultaneously for
all of the orbital elements (10 in the case considered above, of a
double-lined spectroscopic system that is also spatially resolved).
An important advantage of this method is that combined
(three-dimensional) solutions are often possible even if one type of
observation (astrometry or spectroscopy) has insufficient coverage for
an independent fit. In some cases solutions are possible when neither
the astrometric or the spectroscopic observations are sufficient by
themselves. In general these simultaneous fits strengthen the
determination of the elements because they fully exploit the
constraints available from both types of data (redundancy). In
addition, this redundancy allows for useful checks of systematic
errors. Furthermore, the errors in the elements are straightforward to
derive because they come from a single least-squares fit, and do not
suffer from the ambiguities mentioned above (particularly for $P$,
$e$, $\omega$, and $T$). 

Perhaps the only drawback is that combined solutions are perceived by
astronomers to be somewhat more complex mathematically, although this
not necessarily true, as we describe below. 
	
\subsection{The mathematics of combined orbital solutions}

The way in which spectroscopic or astrometric least-squares solutions
are usually posed is well known and conceptually quite simple: the
``best-fit" elements are those that minimize the sum of the normalized
residuals squared, or $\chi^2$. The normalization is achieved by
dividing the residuals by the uncertainty of each measurement, which
provides the weighting. Thus, for an astrometric solution in which the
measurements are the angular separations ($\rho$) and position angles
($\theta$), with uncertainties $\sigma_{\rho}$ and $\sigma_{\theta}$,
respectively, the expression to be minimized is
 \begin{equation}
 \chi^2_{astr} = \sum \left({\rho - \rho^*\over\sigma_{\rho}}\right)^2
+ \sum \left({\theta - \theta^*\over \sigma_{\theta}}\right)^2,
 \end{equation}
 where $\rho^*$ and $\theta^*$ are the predicted values based on the
orbital elements at each iteration, and the sums are carried out over
all observations of each kind. If the measurements are reported in
rectangular rather than polar coordinates, such as those sometimes
made with CCDs or photographic plates, then $x$ and $y$ and their
corresponding errors should be used instead of $\rho$ and $\theta$. 

An analogous expression holds for a spectroscopic solution in which
the velocities $RV_1$ and $RV_2$ of both components are measured,
which is
 \begin{equation}
 \chi^2_{spec} = \sum \left({RV_1 - RV_1^*\over \sigma_{RV_1}}\right)^2
+ \sum \left({RV_2 - RV_2^*\over \sigma_{RV_2}}\right)^2,
 \end{equation}
 where $RV_1^*$ and $RV_2^*$ are again computed at each iteration from
the orbital elements. 

The numerical problem of solving for the orbital elements involves
non-linear equations in both cases, but techniques for tackling such
cases are readily available and are in common use, such as the
Levenberg-Marquardt algorithm (see, e.g., Press et al.\ 1992), and
others. 

In a solution that combines astrometric and spectroscopic
measurements, the figure of merit to be minimized is simply the sum of
the separate $\chi^2$ values, or
 \begin{equation}
\chi^2_{comb} = \chi^2_{astr} + \chi^2_{spec}.
 \end{equation}

As trivial as the above equation may seem, the problem of combining
observations of different kinds to derive a three-dimensional orbit
was first formulated in this way less than 30 years ago by C.\ Morbey
at the Dominion Astrophysical Observatory (Morbey 1975). The author
applied it to the case of the highly eccentric visual binary system
Burnham~1163 (ADS~1123, HD~8556), which had good astrometric coverage
at relatively low precision and spectroscopic coverage only during
periastron, but with much better precision. 

Thus, the mathematics of a combined solution are no more complex than
those of either of the separate solutions, and identical numerical
techniques can be used. 
	
\section{Applications}

One of the most obvious applications of combined
astrometric-spectroscopic solutions is the determination of accurate
stellar masses and orbital parallaxes. The parallaxes along with the
apparent brightness of each component allow the luminosities to be
derived. Masses and luminosities are fundamental data of great
importance for testing models of stellar evolution, particularly in
some regimes where the models are still poorly constrained, such as
the lower main sequence (see, e.g., Henry \& McCarthy 1993; Henry et
al.\ 1999). 

Systems with multiplicity higher than two (triples, quadruples) often
benefit from the simultaneous availability of astrometric and
radial-velocity measurements, and allow the masses of all of the stars
in the system to be determined. These cases are also of interest for
investigating the coplanarity of inner and outer orbits in
hierarchical configurations. 

Another interesting recent application in which spectroscopic and
astrometric information are combined to great advantage is the
determination of masses for substellar objects (brown dwarfs and
extrasolar planets). Such basic information is almost non-existent for
such objects, yet it is crucial to confront the theories that are
being developed for their formation and evolution. 

In the following we describe several specific examples where combined
astr\-om\-etr\-ic-spectroscopic solutions have been carried out, and
we also extend the idea to other types of measurement. 

\subsection{The mass-luminosity relation in the Hyades}

Mass determinations in stellar clusters are particularly interesting
for testing models of stellar evolution because of the additional
constraints that are available: cluster members can all be assumed to
have the same age and chemical composition, and in many cases these
are well known from spectroscopic studies and isochrone fits to the
color-magnitude diagram. The Hyades is a well-known and particularly
important example for this kind of study. Five binary systems in this
cluster have absolute mass determinations, four of which come from
combined astrometric-spectroscopic solutions (51~Tau, 70~Tau,
$\theta^1$~Tau, and $\theta^2$~Tau; Torres, Stefanik, \& Latham
1997a;b;c). The other system (V818~Tau) is an eclipsing binary. The
observational data used for the former 4 systems include visual
measurements (made with filar micrometers or eyepiece
interferometers), measurements by speckle interferometry, lunar
occultation, and long-baseline interferometry, as well as radial
velocity measurements for one or both stars in the binary. The angular
sizes of the orbits range from 19 milli-arcseconds ($\theta^2$~Tau, $P
= 140.7$ days) to $0\farcs13$ ($\theta^1$~Tau, $P = 16.3$ yr). In the
case of $\theta^1$~Tau, which is only a single-lined spectroscopic
binary (the primary star is a giant and the secondary is a
main-sequence star), the distance was used in order to derive the
component masses in the combined solution. For $\theta^2$~Tau the
spectroscopic elements were combined directly with the elements
derived from long-baseline interferometry, since the original
astrometric observations were not available. 

The absolute masses of the 9 main-sequence stars in these binary
systems were used to construct the empirical mass-luminosity relation
in the Hyades, and to compare it with model isochrones computed
specifically for the age and composition of the cluster. Subtle
differences were found, possibly indicating a helium abundance
different from solar. In addition, the orbital parallaxes derived for
51~Tau, 70~Tau, and $\theta^2$~Tau, slightly more precise than those
from the Hipparcos mission, served as a stringent test for systematics
in the satellite determinations and are mentioned on the Hipparcos web
site as a valuable external check (see also de Bruijne, Hoogerwerf,
\& de Zeeuw 2001). 
	
\subsection{Systems of higher multiplicity}

An interesting example of the power of combined
astrometric-spectroscopic solutions in higher multiplicity
configurations is given by the quadruple system $\mu$~Ori, studied by
Fekel et al.\ (2002). This is a visual binary with a highly eccentric
orbit and a semimajor axis of $0\farcs27$ ($P = 18.6$ yr), in which
one component is itself a double-lined spectroscopic binary with a
period of $P = 4.78$ days and the other component is also a
spectroscopic binary (but only single-lined) with a similar period of
$P = 4.45$ days. Visual and speckle observations of the visual pair
were combined with radial-velocity measurements of the 3 visible
objects, and the authors were able to derive the absolute masses of
the stars in the double-lined binary with relative errors of only 2\%.
An orbital parallax was also derived for the system, which is more
than 4 times more precise than the Hipparcos parallax, but is in good
agreement with it. 
	
\subsection{Masses for substellar objects}

In recent years dozens of unseen companions to solar-type stars have
been detected by means of highly-precise radial velocity measurements.
From their small minimum masses these companions appear to be
planetary in nature, and the subject has attracted great attention not
only for studies of the origin and evolution of our solar system, but
also because of its implications for the possibility of
extraterrestrial life.  But since Doppler spectroscopy only provides a
lower limit to the mass of such objects, bringing in complementary
astrometric information is an important application of the techniques
described above. The case of Gl~876 is a good illustration. It is an
M4 dwarf star with two presumably substellar companions in orbit with
periods of about 30 days and 60 days. The star was observed by
Benedict et al.\ (2002) with the Fine Guidance Sensors aboard the
Hubble Space Telescope, and the wobble of the star due to the outer
companion was detected. The authors combined their astrometric
observations with existing radial velocities and solved for the
parallax, proper motion, and the orbital elements of the relative
orbit, which has a semimajor axis of only 0.25 $\pm$ 0.06 mas. Their
determination of the inclination angle ($84\deg \pm 6\deg$) allowed
them to establish the absolute mass of the orbiting companion at 1.89
$\pm$ 0.34 times the mass of Jupiter, thus showing conclusively that
it is a planet. This is the first astrometrically determined mass of
an extrasolar planet. 

Among the more than 100 low-mass companions detected by the Doppler
planet searches, a few of the more massive ones with minimum masses
between about 10 and 65 Jupiter masses were investigated by Halbwachs
et al.\ (2000) to attempt to determine their true masses by
establishing the inclination angles astrometrically. If the masses
were to come out lower than 80 Jupiter masses (the substellar limit),
they would fall in the ``brown dwarf desert", a term used to refer to
the apparent lack of brown dwarf companions with relatively short
periods in Doppler planet surveys of solar type stars.  Hipparcos
intermediate data (abscissa residuals) were combined with the
spectroscopic orbital elements for 11 candidate brown dwarfs, but most
of them were shown to have true masses above the substellar limit, or
only slightly below but with relatively low confidence.  Similar
solutions were attempted later by other investigators to determine the
inclination angles (and therefore the masses) of even lower-mass
(planetary) companions, but those results were shown by Pourbaix
(2001), Zucker \& Mazeh (2001), and Pourbaix \& Arenou (2001) to be
spurious due to numerical reasons (see also the article by D.\
Pourbaix in this volume). 
	
\section{Combined solutions of other kinds}

In the applications described above spectroscopic information was
combined with astrometric measurements, although the latter are
one-dimensional abscissa residuals provided by the Hipparcos mission
(in milli-arcseconds, along a reference great circle), rather than
traditional two-dimensional measurements such as \{$\rho$, $\theta$\}
or \{$x$, $y$\}. Operationally, however, the procedure is similar to
the astrometric-spectroscopic solutions described earlier.  In fact,
the idea of combining measurements of different types is not limited
to astrometry$+$spectroscopy. As an example, spectroscopic
observations can be combined with eclipse timings measured for
eclipsing binaries, based on photometry. The next section illustrates
this with some examples. 
	
\subsection{Solutions using times of eclipse}

The orbital periods of eclipsing binaries are often determined from
the accurate measurement of times of minimum light (for the primary
and/or secondary eclipse). The period and epoch are usually then fixed
for the spectroscopic solution. There are situations, however, where a
combined fit that incorporates the eclipse timings together with the
radial velocities has significant advantages, such as when the times
of minimum and the velocities are separated by a significant interval
of time. In that case combining them gives a much longer time
baseline. Mathematically the eclipse timings are incorporated into the
solution by simply adding the $\chi^2$ term
 \begin{equation}
 \chi^2_{time} = \sum \left({T_I - T_I^*\over \sigma_{T_I}}\right)^2 +
\sum \left({T_{II} - T_{II}^*\over \sigma_{T_{II}}}\right)^2,
 \end{equation} 
 to the spectroscopic term in eq.(2), so that the total $\chi^2$ is
$\chi^2_{comb} = \chi^2_{spec} + \chi^2_{time}$. The symbols $T_I^*$
and $T_{II}^*$ above are the predicted times of eclipse for the
primary and secondary, based on the orbital elements at each
iteration. Examples of binaries where this has been applied are FS~Mon
(Lacy et al.\ 2000) and EI~Cep (Torres et al.\ 2000a). 

A number of eclipsing binaries present apsidal motion (a change in
$\omega$), which over time alters the shape of the radial velocity
curves. Such cases provide another application where the combination
of eclipse timings and velocities into a single least-squares solution
can significantly strengthen the determination of the apsidal motion
constant (d$\omega$/dt). This quantity contains information on the
internal structure of stars (specifically, their degree of mass
concentration), and as such it can provide important constraints to
the theory of stellar interiors.  V364~Lac (Torres et al.\ 1999) and
GG~Ori (Torres et al.\ 2000b) are two cases where this approach has
been followed. 

The types of combined solutions that have been mentioned so far merge
together astrometry and radial velocity information, or astrometry and
eclipse timings. Although one might not think such a case would ever
arise in practice, the combination of astrometry and eclipse timings
is also possible in the system R~CMa, studied by Ribas, Arenou, \&
Guinan (2002). This is an Algol-type eclipsing binary ($P = 1.14$
days) where the residuals from the recorded times of eclipse that span
more than a century show a significant modulation due to the
light-travel time effect.  Additionally, Hipparcos measurements reveal
small acceleration terms (non-linear proper motions). Both of these
effects are nicely explained by the presence of a third star in a
distant orbit with a period of about 93 yr. By incorporating also
ground-based astrometric measurements to constrain the proper motion,
the parameters of the outer orbit can be derived through a global
solution, and its orientation turns out to be consistent with being
coplanar with the inner orbit. 
	
\subsection{Alternative astrometric-spectroscopic solutions}

Spectroscopic observations can be incorporated into combined solutions
in ways other than by using the radial velocities. Forveille et al.\
(1999) applied this technique to the astrometric-spectroscopic binary
Gl~570BC, a pair of M dwarfs resolved by adaptive optics measurements
as well as by 1-D and 2-D speckle interferometry. Instead of using
velocities, they incorporated the cross-correlation profiles directly
into the solution, adding to the usual orbital elements 4 new
parameters that describe the Gaussian correlation peaks for the two
components. Significant numerical advantages and improved errors are
claimed with this technique, as well as less sensitivity to systematic
errors in the velocity amplitudes. The main disadvantage is that there
is no radial-velocity curve to admire, because radial velocities are
bypassed altogether.  Additional examples of this approach include
Gl~234, Gl~747, Gl~831, and Gl~866 (S\'egransan et al.\ 2000). 

Finally, astrometric information can also be incorporated into
combined solutions without directly involving positional measurements
on the sky such as \{$\rho$, $\theta$\}. This is the case, for
example, in long-baseline interferometry where the measurement is the
``fringe visibility", which represents the contrast of the
interference fringes. Visibilities (often used squared, as $V^2$)
implicitly contain information on the relative position of the
components (see, e.g., Boden et al.\ 2000), and can be treated as any
other observable by writing the corresponding $\chi^2$ term
 \begin{equation}
 \chi^2_{astr} = \sum \left({V^2 - (V^2)^*\over \sigma_{V^2}}\right)^2
 \end{equation}
 and combining it with the spectroscopic term in eq.(2). This has a
number of important advantages over the alternative approach, which
would be to derive \{$\rho$, $\theta$\} based on the fringe
measurements on each night. The latter is not always possible if there
are only a few visibilities on a given night, whereas the individual
visibilities can be incorporated directly as indicated in eq.(5) even
if there is only one measurement on that night. In addition, using the
visibilities directly allows one to account for motion even within a
night for short-period systems. This technique has been applied to
Capella (Hummel et al.\ 1994), 12~Boo (Boden et al.\ 2000), HD~195987
(Torres et al.\ 2002), and many other binary systems. 
	
\section{Conclusions}

Simultaneous astrometric-spectroscopic solutions usually have
important advantages compared to separate fits to the astrometry or
the radial velocities. These types of combined fits are still not seen
very often in the literature, but with advances in observational
techniques and increased access to data, they should become more
common. The same approach can be extended to many other types of
observations aside from \{$\rho$, $\theta$\} and velocities, such as
Hipparcos intermediate data (abscisa residuals or transit data), lunar
occultations, interferometric visibilities and closure phases,
parallaxes, positions and proper motions, cross-correlation profiles,
times of eclipse, and even magnitudes. 
	
\acknowledgements

Partial support for this work was provided by NASA's MASSIF SIM Key
project (JPL grant 1240033).

\end{document}